\documentclass[11pt,a4paper]{article}
\usepackage[T1]{fontenc}
\usepackage[margin=2.5cm]{geometry}
\usepackage{amsmath,amssymb,amsthm,mathtools}
\usepackage{enumitem}
\usepackage{hyperref}
\usepackage{cleveref}

\theoremstyle{plain}
\newtheorem{theorem}{Theorem}[section]
\newtheorem{lemma}[theorem]{Lemma}
\newtheorem{example}[theorem]{Example}

\DeclareMathOperator{\rank}{rank}
\DeclareMathOperator{\select}{select}
\DeclareMathOperator{\access}{access}
\DeclareMathOperator{\occ}{occ}

\title{Asymptotically Optimal Representation of Tree-Like Graphs}
\author{Johannes Fischer}

\begin{document}
\maketitle

\begin{abstract}
We show how to represent a graph with $n$ nodes, $m=n+k=O(n)$ edges, and $h\le k$ nodes with more than one incoming edge in asymptotically optimal space, in a way that supports basic navigational operations quickly.

\end{abstract}

\section{Introduction}
\label{sec:intro}
We focus on the representation of graphs with $n$ nodes that are \emph{tree-like}, by which we mean they have $m=O(n)$ edges.
Our parameters of interest are the number $k$ of non-tree edges, and the number $h$ of nodes with more than one incoming node.
Parameter $k$ is also known as the \emph{cyclomatic number} of a graph, and $h$ as its \emph{reticulation number}.
Fischer and Peters \cite{FP16} presented a data structure for such graphs called \emph{GLOUDS} (graph level order unuary degree sequence).
However, they did not show the optimality in the worst case, i.e. showing that the data structure never uses asymptotically more space than necessary for this class of graphs.
We close this gap in this article by showing that (a slightly modified form of) GLOUDS needs at most $(1+\epsilon)\log |G(n,k,h)| + O(n)$ bits of space, where $G(n,k,h)$ is the set of all unlabeld directed graphs with the parameters above.
It supports listing the incoming edges of any node in $O(1)$ time per incoming edge, and the outgoing neighbors in $O(1/\epsilon)$ time.

\subsection{Other Related Work}
Gurevich et al.\ \cite{GSV84} defined $k$-almost trees, where $k$ is counted per binconnected component; this is probably the closest definition to what we call tree-like graphs.
Very recently Ismaili Alaoui et al.\ \cite{INNW26} built upon GLOUDS by compressing some components, and showed space optimality under the \emph{node-degree entropy}.
This is very much different from our approach here, as they try to achieve optimality for a given distribution of in-degrees, important, e.g., when some nodes attract neighbors more likely than others, but saying nothing about optimality over the entire graph class.

The literature on succinct tree representations is rich.
We refer to a recent survey by Chakraborty and Sadakane \cite{chakraborty2025succinct}.

\section{Definitions}
\label{sec:defs}
Let $G=(V,E)$ be a directed graph with $n=|V|$ nodes and $m=|E|$ edges.

A set $R\subseteq V$ is a \emph{root set} of $G$ if every node of $G$ is reachable from some node in $R$. The \emph{number of roots} $c=c(G)$ is the minimum size of a root set.
A minimum root set contains exactly one node of each strongly connected component without incoming edges from other strongly connected components, and $c$ can be computed in linear time with Tarjan's algorithm~\cite{Tar72}.
Let $R=\{r_1,\dots,r_c\}$ be a minimum root set.
A \emph{spanning forest} of $G$ is a set $F\subseteq E$ such that every $v\in V\setminus R$ has exactly one incoming edge in $F$, no root has an incoming edge in $F$, and every node is reachable from a root via edges of~$F$.
The edges in $E\setminus F$ are the \emph{non-tree edges}; their destinations form the set $H$ of \emph{non-tree nodes}.
We write $k:=|E\setminus F|$ and $h:=|H|$.
Note that $m=n+k-c$, and $v\in H$ iff either $v$ has in-degree at least 2, or $v\in R$ and $v$ has in-degree $\ne 0$.

We use the following results \cite{RRR07} on succinct data structures for rank/select (in a bit-vector, rank counts the number of $1$s or $0$s up to a given position $i$, and select gives the position of the $i$'th $1$s or $0$s):

\begin{lemma}
\label{lem:bv}
\begin{enumerate}[label=(\alph*)]
  \item (\emph{Fully indexable dictionary}) A bit vector of length $N$ with $K$ ones can be stored in $\lg\binom NK+O(N\lg\lg N/\lg N)$ bits such that rank and select (for $0$s and $1$s) take $O(1)$ time.
  \item (\emph{Indexable dictionary}) A set $S\subseteq[0,N)$ with $|S|=K$ can be stored in $\lg\binom NK+o(K)+O(\lg\lg N)$ bits such that select, membership, and the rank of an element $x\in S$ (called \emph{partial rank}) take $O(1)$ time.
\end{enumerate}
\end{lemma}

A sequence just needing select and access can be represented like this (see, e.g. \cite[p127]{Nav16}):

\begin{lemma}
\label{lem:seq}
Let $S[0,k)$ be a sequence over the alphabet $[0,h)$, and let $t\ge1$.
Then $S$ can be stored in $k\lg h+O(\frac kt\lg h)+O(k)$ bits such that $\select_a(S,j)$ (the position of the $j$'th $a$ in $S$) and the number of occurrences $\occ_a(S)$ of a symbol take $O(1)$ time, and $\access(S,i)=S[i]$ takes $O(t)$ time.
\end{lemma}

\section{Recap of GLOUDS}
\label{sec:recap}
Here, we briefly re-prove the following theorem, which combines ideas from \cite{FP16,INNW26} and uses different auxilliary data structures for rank/select. Although the precise statement has not appeared before, we do not claim that in contains significant new ideas; the purpose is just completeness and a form that enables to prove our optimality claim.

\begin{theorem}
\label{thm:space}
Let $G$ be a directed graph with $n$ nodes, $m$ edges, $c$ roots, $k=m-n+c$ non-tree edges and $h$ non-tree nodes, and let $t\ge1$. Then $G$ can be represented in
\[
  2n+\log\binom{n+k}{k}+\log\binom nh+k\log h+O(k+\frac{k}{t}\log h)+o(n)
\]
bits such that the $y$ out-neighbours of a node are listed in $O(1+t\,y)$ time, the $x$ in-neighbours in $O(1+x)$ time, and the in- and out-degree are computed in $O(1)$ time.
\end{theorem}
\begin{proof}
We compute a \emph{spanning forest} of $G$ with roots $R$, and connect all $r\in R$ to an artificial super-root, resulting in a spanning tree $T$ of the augmented graph.
$T$ is represented with LOUDS \cite{Jac89}, using $2n+o(n)$ bits for rank and select (Lemma \ref{lem:bv}(a)).
Define $D=\texttt{1}^{s_1}\texttt{0}\,\texttt{1}^{s_2}\texttt{0}\cdots\texttt{1}^{s_n}\texttt{0}$, where $s_v$ is the number of children of $v$ not already in $T$ (called \emph{shadow children} in \cite{FP16}).
$D$ has length $n+k$ and contains $k$ ones, and we store it in a fully indexable dictionary (Lemma \ref{lem:bv}(a)) using $\lg\binom{n+k}k+o(n+k)$ bits.
This represents the number of non-tree children per node.
For representing the non-tree nodes, define a bit vector $C[0,n]$ with $C[v]=1$ iff $v\in H$ (and $C[0]=0$ for the super-root), stored as an indexable dictionary (Lemma \ref{lem:bv}(b)) in $\lg\binom nh+o(h)+O(\lg\lg n)$ bits.
Finally, we need to represent where the non-tree edges point to.
To this end, we use a bit vector $H[0,k)$ with $H[j]=\rank_1(C,w)\in[0,h)$, where $w$ is the destination of the $j$'th non-tree edge.
Non-tree edges are first ordered by the level order of their origin, and then by destination.
$H$ is stored with Lemma \ref{lem:seq}.
Summing over the the space for $T$, $D$, $C$, and $H$ gives the space bound.
Note that the representation of the non-tree edges is closer to \cite{INNW26} than to \cite{FP16}, but we do not focus on compressing the tree entropy.

A node $v\in[1,n]$ corresponds to the $v$'th $1$ in $T$.
The shadow children of $v$ are $H[a_v,a_v+s_v)$, where
  \[
    a_v=\select_0(D,v-1)-v+2\text{~and~}s_v=\select_0(D,v)-\select_0(D,v-1)-1 .
  \]
Computing in- and out-neighbors is exactly as described in \cite{FP16}.
The out-degree is computed as $\select_0(T,v+1)-\select_0(T,v)-1+s_v$.
Likewise, the in-degree is $[\text{parent}_T(v)\ne 0]+C[v]\cdot\occ_{\rank_1(C,v)}(H)$.
\end{proof}

\begin{example}
\label{ex:fig2}
For the example in \cite[Fig.~2]{FP16}, we have:
$T=\texttt{10\,110\,10\,10\,110\,10\,0\,0\,0}$,
$D=\texttt{0\,0\,10\,0\,10\,10\,0\,10}$,
$C=\texttt{0\,00010011}$, and
$H=(0,1,2,1)$.
Some operations are:
$\mathrm{parents}(7)$: the tree parent is $4$; $a=\rank_1(C,7)=1$; the symbol $1$ occurs at positions $1$ and $3$ of $H$, and $\select_1(D,2)=5$, $\select_1(D,4)=10$ give the tails $\rank_0(D,5)+1=5$ and $\rank_0(D,10)+1=8$.
Similarly, $\mathrm{children}(5)$: tree child $8$; $a_5=1$, $s_5=1$, $H[1]=1$ gives the shadow child $\select_1(C,2)=7$.
\end{example}

\section{Optimality}
\label{sec:optimality}

To show that GLOUDS achieves asymptotically optimal space, we define a sufficiently large subclass of all graphs and bound the size of this class from below.

\subsection{A Large Enough Class of Graphs}
To this end, let $G(n,k,h)$ denote the set of directed graphs on $n$ nodes and $m$ edges, where each $G\in G(n,k,h)$ satisfies the following constraints: 
\begin{enumerate}
  \item Exactly one node $r$ has in-degree 0, and every other node is reachable from $r$.
  \item $G$ has $k$ non-tree edges, hence $m=n-1+k$.
  \item $G$ has $h$ non-tree nodes (nodes with in-degree $>1$).
  \item There are no parallel edges or loops.
\end{enumerate}
A first difficulty we face is that the graphs in $G(n,k,h)$ are unlabeled, such that there are many representations of $G \in G(n,k,h)$ in any data structure that (implicitly) labels the graphs.
E.g., in GLOUDS we have at least two degrees of freedom: the choice of the spanning tree, and the order of the non-tree nodes in $H$, each choice resulting in a potentially different labeling of the nodes.
We tackle this by letting $G_l(n,k,h)$ denote the set of \emph{labeled} graphs with the same constraints as above.
Note that $|G(n,k,h)| \ge \frac{|G_l(n,k,h)|}{n!}$, since there are $n!$ node labelings.

For bounding the size of $G_l(n,k,h)$ from below, we proceed by counting the number of ways for generating the graphs in $G_l(n,k,h)$ from labeled spanning trees, i.e.,
by counting the number of pairs $(T,G)$ with $G \in G_l(n,k,h)$ and $T$ being a spanning tree of $G$.
From this, we derive a lower bound for $G_l(n,k,h)$ by dividing out the number of choices for the spanning tree.

So let $P$ be the set of pairs $(G,T)$ with $G=(V,E) \in G_l(n,k,h)$ and $T$  a spanning tree of $G$.
We count the size of $P$ by starting from spanning tree, then identifying $h$ non-tree nodes, and finally identifying non-tree edges.
Each step counts the number of ways this can be done; the product gives $|P|$.

We start with the number of spanning trees.
By Cayley's formula, there are $n^{n-2}$ different labeled unrooted trees on $n$ nodes.
As any of its nodes can be chosen as a root, there are $n^{n-1}$ different labeled rooted spanning trees.

With a fixed spanning tree $T$, we now look at how to generate a graph from $T$ with the desired properties.
From the $n-1$ nodes other than the root, we can choose $h$ non-tree nodes in $\binom{n-1}{h}$ different ways.

Finally, we move to the choices of non-tree \emph{edges}.
Define $f_{h,n}(k)$ as the number of ways to construct a graph $G\in G_l(n,k,h)$ from a fixed spanning tree, where $h\le n$ nodes have already been chosen as non-tree nodes.
Note that with this definition we have
\begin{equation}
\label{eqn:p}
|P| = n^{n-1} \times \binom{n-1}{h} \times f_{h,n}(k)
\end{equation}

For any of the $h$ non-tree nodes $v$ we have to choose a set of $d(v)$ extra parents, from any of $n-2$ nodes except $v$ and its tree-parent; there are $\binom{n-2}{d(v)}$ many.
Take a fixed distribution $(d_1 \ge 1, \dots, d_h\ge 1)$ of the number of extra parents for the $h$ non-tree nodes.
Note that $\sum d_i = k$.
For the whole tree there are $\prod_{1\le i \le h} \binom{n-2}{d_i}$ possibilities.
Summing over all possible distributions of extra parents, we see that the total number of possibilities is
$$
f_{h,n}(k) = \sum_{(d_1,\dots, d_h) \in [1,k]^h,  \sum d_i = k}\prod_{1\le i \le h} \binom{n-2}{d_i}
$$

For deriving a lower bound on $f_{h,n}(k)$,
we look at how a graph with $i\ge h$ extra nodes attaching to the $h$ non-tree nodes can be extended to one with $i+1$ non-tree nodes, and set up a recursion.
When $i=h$, for each non-tree node we can choose $n-2$ nodes as their extra parent (exluding the node itself and its parent).
Hence, $f_{h,n}(h) = (n-2)^h$.
For the recursive case $k\ge i>h$, we can add one of the $h(n-2)-i+1$ remaining possible edges to any of the $f'_{h,n}(i-1)$ configurations having one edge less, hence with $M:=h(n-2)$ we have

$$
f_{h,n}(i) = (M-i+1)f_{h,n}(i-1)\ .
$$

Unrolling the recursion and evaluating $f$ at $k$ gives (with $N:=n-2$):

$$
f_{h,n}(k) = N^h \prod_{j=h}^{k-1} (M-j) = N^h\frac{(M-h)!}{(M-k)!}\ .
$$

Setting this in relation to the quantity $\binom{M}{k}$ (which is the number of choices of non-tree nodes \emph{ignoring} the fact that any non-tree node needs at least one extra parent), we get:

\begin{eqnarray*}
\frac{f_{h,n}(k)}{\binom{M}{k}} & = & N^h\frac{(M-h)!/(M-k)!}{M!/(k!(M-k)!)}\\
 &   = & \frac{N^hk!}{\prod_{i=1}^{h-1}(M-i)}\\
 & \ge & \frac{N^hk!}{M^h} = \frac{N^hk!}{h^hN^h}= \frac{k!}{h^h}\\
 & \ge & \frac{h!}{h^h} = e^{\log h! - h\log h} \ge e^{h\log h - h+1 - h\log h} = e^{1-h} \ge e^{-h}\ ,
\end{eqnarray*}
hence $f_{h,n}(k) \ge \binom{M}{k}e^{-h}$.

With \eqref{eqn:p}, this gives us
\begin{equation}
\label{eq:plb}
|P| \ge n^{n-1} \times \binom{n-1}{h} \times \binom{h(n-2)}{k}e^{-h}\ .
\end{equation}

We now wish to relate $|P|$ with $|G_l(n,k,h)|$.
Recall that in $P$ we started by counting \emph{all} labeled spanning trees.
But the same labeled graph can be represented with several different spanning trees, and we want to count it only once.
To correct for this, take a fixed $G \in G_l(n,k,h)$, let $S$ be the set of its $h$ non-tree nodes, and let again $d(v)$ denote the number of non-tree parents for $v\in S$.
Let $\tau(G)$ denote the number of $G$'s possible spanning trees.
Each spanning tree has exactly one parent per node in $S$ (though not all choices yield a spanning tree), so
\begin{eqnarray*}
\tau(G) & \le & \prod_{v\in S}(1+d(v))\\
 & \le & (\frac{1}{h}\sum_{v\in S}(1+d(v)))^h\\ 
 & =   & (\frac{1}{h}(h+\sum_{v\in S}d(v))^h\\
 & =   & (1+\frac{k}{h})^h.
\end{eqnarray*}
By summing over all labeled graphs, we get

$$
|P| = \sum_{G \in G_l(n,k,h)}\tau(G) \le |G_l(n,k,h)| \times (1+\frac{k}{h})^h\ ,\text{~so}
$$ 
\begin{equation}
\label{eq:glb}
|G_l(n,k,h)| \ge \frac{|P|}{(1+\frac{k}{h})^h}\ .
\end{equation}

Combining \eqref{eq:plb} with \eqref{eq:plb}, we finally get
\begin{eqnarray*}
|G(n,k,h)| & \ge & \frac{|G_l(n,k,h)|}{n!}\\
 & \ge & \frac{|P|}{n!(1+\frac{k}{h})^h}\\
 & \ge & \frac{n^{n-1}}{n!} \times \binom{n-1}{h} \times \binom{h(n-2)}{k}e^{-h} \times (1+\frac{k}{h})^{-h}\\
 & \ge & e^{n-1}n^{-3/2} \times \binom{h(n-2)}{k}e^{-h} \times (1+\frac{k}{h})^{-h}\ .
\end{eqnarray*}

Taking logarithms, we get the following lemma:
\begin{lemma}
\label{lemma:lb}
A representation of a simple graph with exactly one root, $h\le \min(n-1, k)$ non-tree nodes, and $h(n-2) \ge k\ge h$ non-tree edges needs at least
$$
\log\binom{n-1}{h} + \log\binom{h(n-2)}{k} +(n-1)\log e -\frac{3}{2}\log n - h\log(e(1+\frac{k}{h})) 
$$
bits.
\end{lemma}

\subsection{Final Bound}
Since GLOUDS represents (among others) the graphs in Lemma \ref{lemma:lb}, we can use it to show the following:

\begin{theorem}
For every $t\ge 1$, GLOUDS represents a graph from $G(n,k,h)$ (graphs with $n$ nodes, $h\le n$ non-tree nodes and $k$ non-tree edges) in $(1+O(1/t)) \log |G(n,k,h)| + O(n)$ bits, given that $k=O(n)$, i.e., for tree-like graphs.
\end{theorem}
\begin{proof}
We just focus on the case $k>0$. Simplifying Lemma \ref{lemma:lb} with $k=O(n)$, we know that we need at least 
\begin{equation}
\label{eq:simplelb}
\log |G(n,k,h)| \ge \log\binom{n-1}{h} + \log\binom{h(n-2)}{k}-O(n)
\end{equation}
bits, whereas the space of Theorem \ref{thm:space} is $S+O(\frac{k}{t}\log h)$ with

\begin{equation}
\label{eq:simplespace}
     S:=\log\binom{n+k}{k}+k\log h+\log\binom nh + O(n)
 \le  k\log h+\log\binom nh + O(n)
\end{equation}

bits, the inequality following from $\binom{n+k}{k} \le 2^k$.
We want to set \eqref{eq:simplelb} and \eqref{eq:simplespace} in relation.
Dropping the $O(n)$ terms and starting from \eqref{eq:simplelb}, we have:

\begin{eqnarray*}
\log\binom{n-1}{h} + \log\binom{h(n-2)}{k} & \ge & \log (n^{-1} \binom{n}{h}) + \log\binom{h(n-2)}{k}\text{~(using~}h \le n-1\text{)}\\
& \ge & \log (n^{-1} \binom{n}{h}) + k\log h - k\log\frac{k}{n-2}\\
& \ge & \log \binom{n}{h} + k\log h - O(n)\text{~(using~}k=O(n)\text{)}\\
& \ge & S-O(n)\ ,
\end{eqnarray*}
so the final space for GLOUDS is at most
\begin{eqnarray*}
     & \log\binom{n-1}{h} + \log\binom{h(n-2)}{k} +O(\frac{k}{t}\log h) + O(n)\\
 \le &  (1+O(1/t))(\log\binom{n-1}{h} + \log\binom{h(n-2)}{k}) + O(n) \\
   = &  (1+O(1/t))\log |G(n,k,h)| + O(n) 
\end{eqnarray*}
bits, the inequality following from \eqref{eq:simplespace}.
\end{proof}
We note that with some more effort it could be shown that for some combinations of $n,k,h$ GLOUDS is \emph{succinct}, meaning that the $O(n)$ term is indeed $o(|G(n,k,h)|)$.
Showing this in general is left for further research.

\bibliographystyle{alpha}
\bibliography{literatur}

\end{document}